\documentclass[twocolumn,english,twocolumn,english,aps,prb,floats,showpacs]{revtex4}
\usepackage[T1]{fontenc}
\usepackage[latin9]{inputenc}
\usepackage{amsmath}
\usepackage{amssymb}
\usepackage{graphicx}
\usepackage{esint}

\makeatletter
\@ifundefined{textcolor}{}
{%
 \definecolor{BLACK}{gray}{0}
 \definecolor{WHITE}{gray}{1}
 \definecolor{RED}{rgb}{1,0,0}
 \definecolor{GREEN}{rgb}{0,1,0}
 \definecolor{BLUE}{rgb}{0,0,1}
 \definecolor{CYAN}{cmyk}{1,0,0,0}
 \definecolor{MAGENTA}{cmyk}{0,1,0,0}
 \definecolor{YELLOW}{cmyk}{0,0,1,0}
}

\usepackage{babel}

\usepackage{babel}

\makeatother

\usepackage{babel}
\begin{document}

\title{Thermal fluctuations of the Josephson current in a ring of superconducting
grains}

\author{D. A. Garanin and E. M. Chudnovsky}

\affiliation{Physics Department, Lehman College and Graduate School, The City
University of New York, 250 Bedford Park Boulevard West, Bronx, NY
10468-1589, U.S.A.}

\date{\today}
\begin{abstract}
Thermal fluctuations of the Josephson current induced by the magnetic
flux through a ring of $N$ superconducting grains are studied. When
a half-fluxon is threading the ring, $I$ exhibits incoherent transitions
between the two degenerate states due to thermal phase slips. We propose
a new numerical method to deal with both equilibrium and dynamic properties
of Josephson systems. Computed transition rate has the form $\Gamma=A(N)\exp[-B(N)/T]$,
where $B(N)$ agrees with the analytical result derived for the energy
barrier associated with phase slips. In the non-degenerate case (e.g.,
at a quarter-fluxon) the equilibrium value of $I$ decreases with
$T$ due to harmonic excitations and then gets destroyed by phase
slips.
\end{abstract}

\maketitle

\section{Introduction}

Persistent currents in small metallic rings have been studied theoretically
and experimentally since 1960s. \cite{Oreg} Rapid progress in manufacturing
of nanostructures \cite{Mooij-2015} has ignited a contemporary interest
to measurements of microscopic chains of Josephson junctions (JJ).
\cite{Pop} Analytical studies in this area consider two limits: When
the dynamics of the chain is dominated by the capacitances of the
junctions \cite{Matveev} and when the dynamics is dominated by the
capacitances of the superconducting islands. \cite{Choi-1993} Here
we focus on the latter limit in the classical regime when thermal
fluctuations dominate over quantum fluctuations.

There are two characteristic energy scales in the problem. One is
the charging energy of the superconducting island, $U\equiv E_{C}=2e^{2}/C$,
where $C$ is the capacitance of the island with respect to the ground,
and the other is the Josephson energy, $J$. They determine the characteristic
temperature ranges and physical properties of the JJ chains. \cite{Bradley,Choi-1993,Wallin-94,Girvin,Sachdev}
At $T\ll T^{*}=\sqrt{2JU}$ quantum fluctuations dominate over thermal
fluctuations. At $T=0$ and $T^{*}=T_{KT}\sim J$ (with $T_{KT}$
being the temperature of the Kosterlitz-Thouless transition in a $2d$
XY model) quantum phase slips yield the superconductor-insulator transition.
\cite{Zaikin,Golubev,Korshunov,Chow} The persistent currents in the
quantum regime have been computed numerically for long chains, \cite{Lee-2003,garchuPRB16}
as well as analytically using the effective low-energy description.
\cite{Rastelli}

Here we focus on the classical thermal regime corresponding to the
temperature range $T^{*}\ll T$, which is easily accessible in experiment.
We begin with analytical calculation of the low-temperature behavior
of the persistent current $I$ and the energy barrier for the phase
slip. The numerical computation of the equilibrium value and dynamics
of $I$ that follows is challenging for two reasons. Firstly, while
fluctuations of the current decrease with the length of the chain,
so does the current itself, $I\propto1/N$. Thus increasing the system
size does not suppress fluctuations and an extensive averaging is
needed. Secondly, accounting for the exponentially rare phase slips
at $T\ll J$ requires a very long computer time. We propose an efficient
numerical method to compute both equilibrium and dynamic properties.

\section{The model}

The energy of the ring is a sum of charging energies of the grains
with and the Josephson coupling energy (see, e.g. Ref.\ \onlinecite{garchuPRB16}
and references therein)
\begin{equation}
\mathcal{H}=\sum_{i=1}^{N}\left\{ \frac{C}{2}V_{i}^{2}+J\left[1-\cos\left(\theta_{i+1}-\theta_{i}+\frac{2\pi\phi}{N}\right)\right]\right\} .\label{Hamiltonian-initial}
\end{equation}
Here $C$ is the capacitance of a superconducting grain with respect
to the ground, $V_{i}$ is the voltage of the $i$-th grain, $\phi=\Phi/\Phi_{0}$
with $\Phi$ being the magnetic flux piercing the ring and $\Phi_{0}=h/(2e)$
being the flux quantum. Using the Josephson relation
\begin{equation}
V_{i}=\frac{\hbar}{2e}\dot{\theta}_{i},
\end{equation}
where $\theta_{i}$ is the phase of the superconducting order parameter
of the $i$-th grain ($\dot{\theta}_{i}=d\theta_{i}/dt$), one can
rewrite the energy as
\begin{equation}
\mathcal{H}=\sum_{i=1}^{N}\left\{ \frac{\hbar^{2}}{4U}\dot{\theta}_{i}^{2}+J\left[1-\cos\left(\theta_{i+1}-\theta_{i}+\frac{2\pi\phi}{N}\right)\right]\right\} \label{Hamiltonian}
\end{equation}
with $U\equiv E_{C}=2e^{2}/C$. Mechanical analogy to our problem
is a chain of rotators with the moment of inertia $\frac{\hbar^{2}}{2U}$.

Due to periodicity of the ring, the sum rule $\sum_{i=1}^{N}(\theta_{i+1}-\theta_{i})=2\pi m$
with $m$ being an integer, $0\leq m\leq N-1$, is satisfied. The
limitation on $m$ is similar to that on the wave vector in the Brillouin
zone. For a given $m$ the minimum of the Josephson energy is achieved
when all phase differences are the same, $\theta_{i+1}-\theta_{i}=2\pi m/N$:
\begin{equation}
E_{J}^{(m)}=NJ\left[1-\cos\frac{2\pi(m+\phi)}{N}\right].\label{E-min}
\end{equation}
In the half-integer-fluxon case, $\phi=n+1/2$, this ground state
is degenerate, $E_{J}^{(-n)}=E_{J}^{(-n-1)}=NJ\left(1-\cos\frac{\pi}{N}\right)$.
For $0<\phi<1/2$ the ground and first excited states are $m=0,-1$,
their energy difference being
\begin{equation}
E_{J}^{(-1)}-E_{J}^{(0)}=2NJ\sin\frac{\pi}{N}\sin\frac{\pi(1-2\phi)}{N}.
\end{equation}
 This and all other energy differences become small for large $N$.

\begin{figure}
\begin{centering}
\includegraphics[width=8cm]{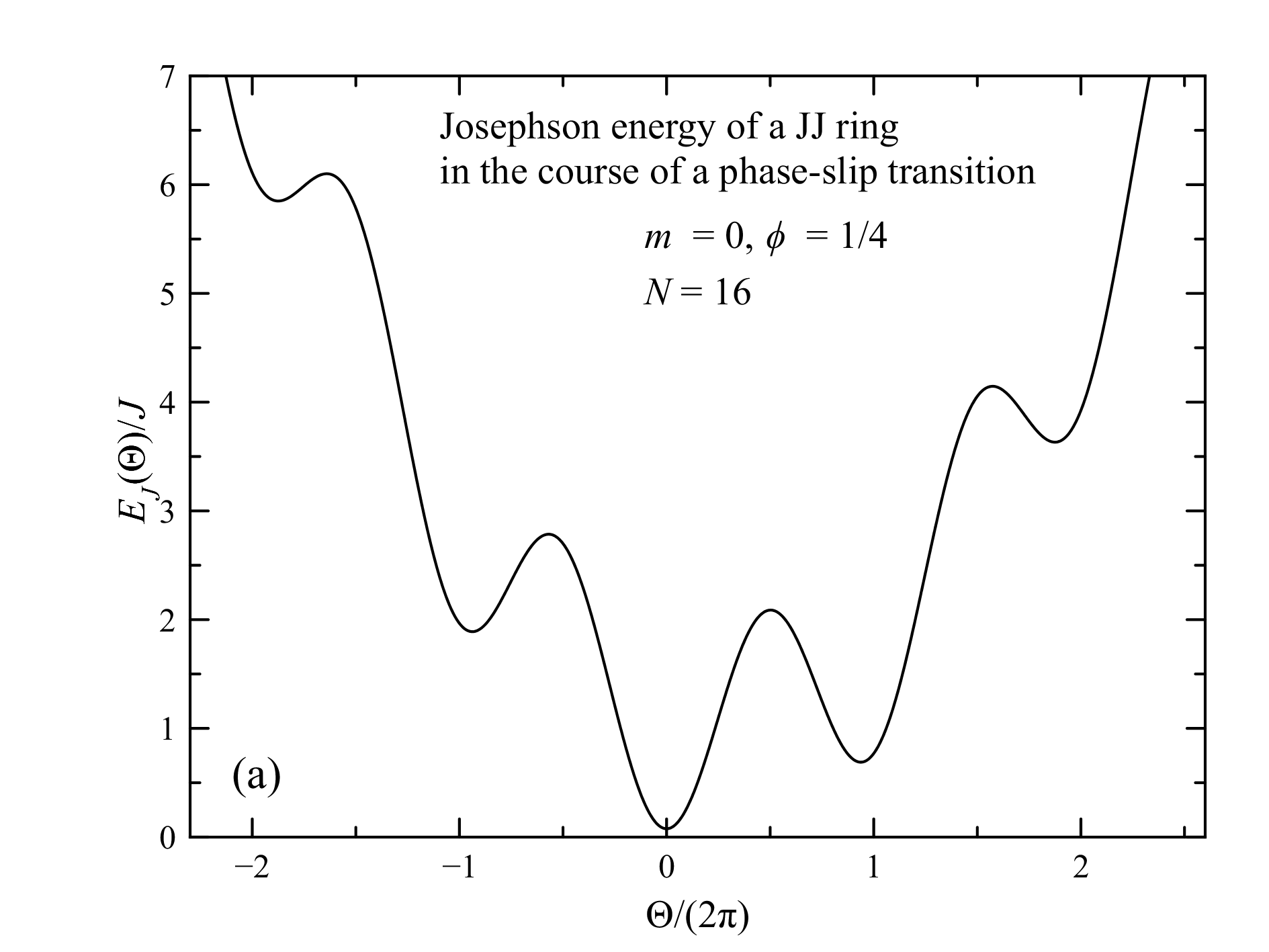}
\par\end{centering}
\begin{centering}
\includegraphics[width=8cm]{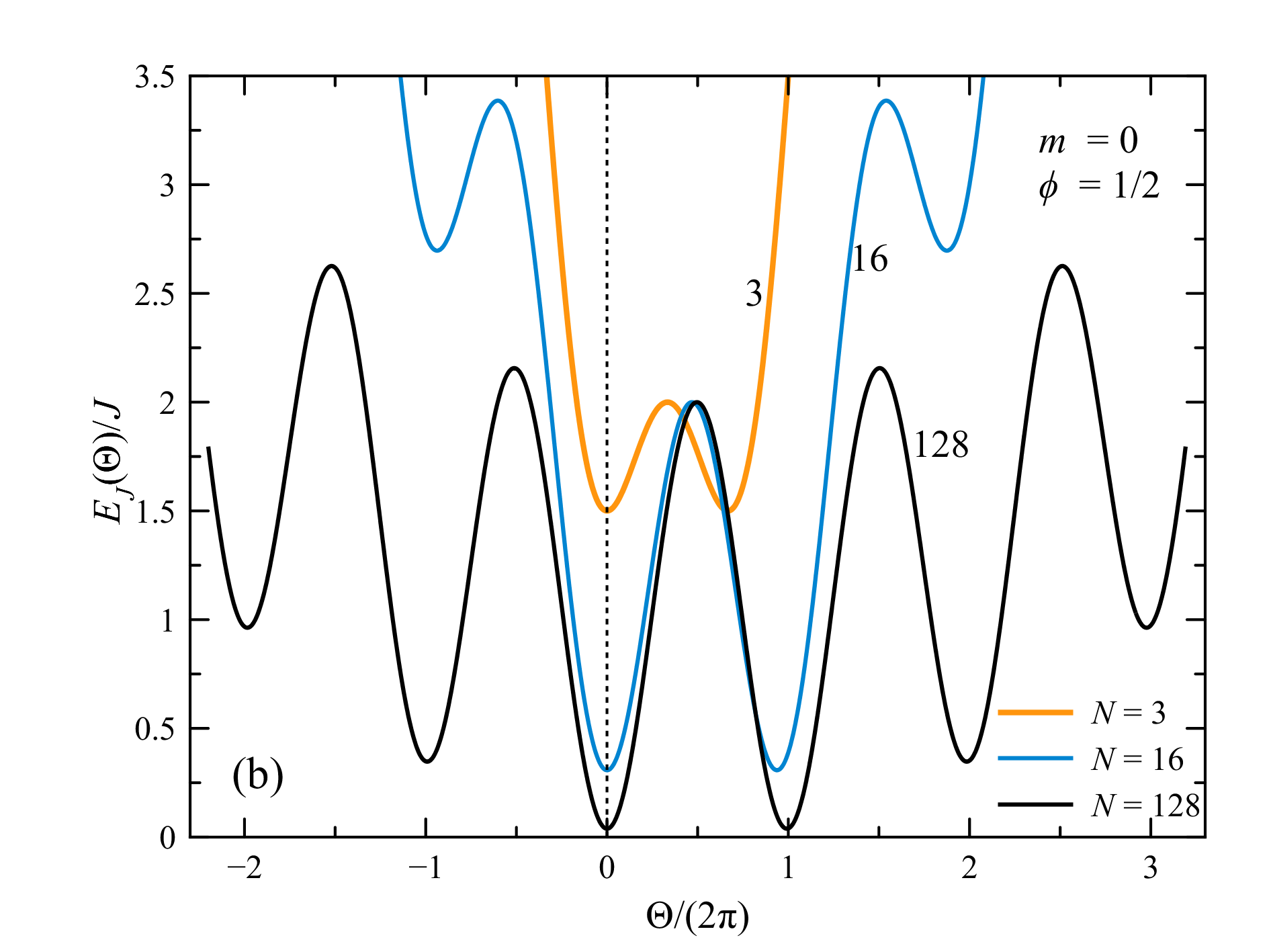}
\par\end{centering}
\caption{Josephson energy $E_{J}$ vs the phase-slip angle $\Theta$, Eq.$\,$(\ref{eq:EJ_Theta}).
(a) $\phi=1/4$. (b) $\phi=1/2$.}

\label{Fig_barrier}
\end{figure}

Consider the energy barrier for a phase slip. To change $m$ in a
manner that requires minimum work, one has to change the phase difference
$\Theta$ between any pair of neighboring grains from nearly zero
to nearly $2\pi$, keeping all other phase differences small and constant,
$\theta_{i+1}-\theta_{i}=\Delta\theta$. Eliminating $\Delta\theta$
from the periodicity condition $\Theta+(N-1)\Delta\theta=2\pi m$
yields the Josephson energy \cite{Rastelli}
\begin{eqnarray}
E_{J}(\Theta) & = & NJ-J\cos\left(\Theta+\frac{2\pi\phi}{N}\right)\nonumber \\
 & - & \left(N-1\right)J\cos\left(\frac{2\pi m-\Theta}{N-1}+\frac{2\pi\phi}{N}\right)\label{eq:EJ_Theta}
\end{eqnarray}
shown in Fig.$\,$\ref{Fig_barrier}. For $N\gg1$ the second cosine
becomes a parabola with superimposed oscillations due to the first
cosine. Transition from $m$ to $m'=m-\eta$ with $\eta=\pm1$ occurs
via changing $\Theta$ from $\Theta^{(m)}=\Delta\theta^{(m)}=2\pi m/N$
to $\Theta^{(m')}=\Delta\theta^{(m')}+2\pi\eta$. Here from the periodicity
condition in the form $2\pi\eta+N\Delta\theta^{(m')}=2\pi m$ one
obtains $\Delta\theta^{(m')}=2\pi m'/N$. In particular, transition
from $m=0$ to $m'=-1$ ($\eta=1)$ requires the change in $\Theta$
from zero to $2\pi(1-1/N)$. Analysis of $E_{J}(\Theta)$ shows that
for $\phi=1/2$ it is symmetric with the top of the energy barrier
between the two minima at $\Theta_{b}=\pi(1-1/N)$, so that
\begin{equation}
B=E_{J}(\Theta_{b})-E_{J}(0)=J\left[2-N\left(1-\cos\frac{\pi}{N}\right)\right].\label{barrier}
\end{equation}
The barrier varies from $J/2$ at $N=3$ to $2J$ at $N\rightarrow\infty$.

Classical equation of motion corresponding to Eq.$\,$(\ref{Hamiltonian})
reads
\begin{equation}
\frac{\hbar^{2}}{2U}\ddot{\theta}_{i}=-\frac{\partial\mathcal{H}}{\partial\theta_{i}}.
\end{equation}
In terms of grain charge $Q_{i}=CV_{i}=\frac{e\hbar}{U}\dot{\theta}_{i}$
this becomes continuity equation
\begin{equation}
\dot{Q}_{i}=-\frac{2e}{\hbar}\frac{\partial\mathcal{H}}{\partial\theta_{i}}=-\frac{2\pi}{\Phi_{0}}\frac{\partial\mathcal{H}}{\partial\theta_{i}}=I_{i,i+1}+I_{i,i-1,}
\end{equation}
where
\begin{equation}
I_{i,i\pm1}=\frac{2\pi J}{\Phi_{0}}\sin\left(\theta_{i\pm1}-\theta_{i}+\frac{2\pi\phi}{N}\right)
\end{equation}
is the current flowing into grain $i$ from grain $i\pm1$. For the
chain current in the direction of increasing $i$ we will use the
average

\begin{equation}
I=\frac{2\pi J}{\Phi_{0}}\frac{1}{N}\sum_{i=1}^{N}\sin\left(\theta_{i+1}-\theta_{i}+\frac{2\pi\phi}{N}\right).\label{eq:I_Def}
\end{equation}
This formula also can be obtained as $I=\partial\mathcal{H}/\partial\Phi$.

In terms of the dimensionless momenta $p_{i}$ defined via $\dot{\theta}_{i}=\frac{\sqrt{2JU}}{\hbar}p_{i}$,
the kinetic energy in Eq.$\,$(\ref{Hamiltonian}) becomes $E_{k}=\sum_{i}\frac{J}{2}p_{i}^{2}$,
and with the dimensionless time $\tau=\frac{\sqrt{2JU}}{\hbar}t$
equations of motion become
\begin{equation}
\frac{dp_{i}}{d\tau}=\sin(\theta_{i+1}-\theta_{i})+\sin(\theta_{i-1}-\theta_{i}),\quad\frac{d\theta_{i}}{d\tau}=p_{i}.\label{eq-motion}
\end{equation}
This system is equivalent to a closed chain of interacting rotators,
with the charging energy playing the role of kinetic energy and the
Josephson energy being potential energy. Here we study the limit of
negligible dissipation which does not show up on the time scale of
the experiment. We used Wolfram Mathematica with compilation in C.
As the differential-equation solver we used the 5th order Butcher's
Ruge-Kutta method that makes 6 function evaluations per integration
step. High precision of this integrator allows using a larger integration
step $\Delta\tau=0.2$.

\section{Equilibrium properties}

To consider equilibrium properties analytically at low temperatures,
it is convenient to introduce reduced phases $\tilde{\theta}_{i}$
according to $\theta_{i}=\tilde{\theta}_{i}+\frac{2\pi m}{N}(i-1)$
(so that accumulation of the reduced phases over the ring is zero).
Thermal average of the ring's Josephson energy $E_{J}\equiv\left\langle \mathcal{H}_{J}\right\rangle $
is given by \cite{garchuPRB16}
\begin{eqnarray}
E_{J} & = & NJ\left[1-\left\langle \cos\left(\tilde{\theta}_{i+1}-\tilde{\theta}_{i}+\frac{2\pi(\phi+m)}{N}\right)\right\rangle \right]\nonumber \\
 & = & NJ\left[1-\left\langle \cos\left(\frac{2\pi(\phi+m)}{N}\right)\right\rangle \left\langle \cos\left(\tilde{\theta}_{i+1}-\tilde{\theta}_{i}\right)\right\rangle \right],\nonumber \\
\end{eqnarray}
where we have taken into account $\left\langle \sin\left(\tilde{\theta}_{i+1}-\tilde{\theta}_{i}\right)\right\rangle =0$
and decoupled fluctuations of the winding number $m$ and reduced
phases $\tilde{\theta}_{i}$. The latter describe harmonic fluctuations
that are similar to spin-wave theory for the equivalent system of
the two-component classical spins. Thus one can use the known result
for the $XY$ classical spin chain in one dimension,
\begin{equation}
\langle\cos(\tilde{\theta}_{i+1}-\tilde{\theta}_{i})\rangle=1-\frac{T}{2J}.
\end{equation}
In a similar way, or just by $I=\partial E_{J}/\partial\Phi$, one
obtains
\begin{equation}
I=\frac{2\pi J}{\Phi_{0}}\left\langle \sin\frac{2\pi(\phi+m)}{N}\right\rangle \left(1-\frac{T}{2J}\right).\label{current_harmonic_approx}
\end{equation}

At $T\ll J$ phase slips changing $m$ are exponentially rare, and
one can discard averaging in Eq.$\,$(\ref{current_harmonic_approx}).
For $\phi=0$ the ground state is $m=0$, and the corresponding current
is zero. For $\phi=1/2$, there are two opposite $I$ values in the
degenerate ground states $m=0,-1$. Eq.$\,$(\ref{current_harmonic_approx})
is valid within time intervals between rare phase slips $0\rightleftharpoons-1$.
However, the large-time average of $I$ is zero. To the contrary,
in non-degenerate cases, such as $\phi=1/4$, there is a robust thermal
average value of $I$.

At higher temperatures one has to take into account thermal fluctuations
of $m$ that are especially pronounced at large $N$ since energy
differences between states with different $m$ decrease with $N$
(see Fig. \ref{Fig_barrier}b). Averaging over $m$ can be done by
\begin{equation}
I=\frac{2\pi J}{\Phi_{0}}\left(1-\frac{T}{2J}\right)\frac{1}{Z}\sum_{m=0}^{N-1}\sin\frac{2\pi(\phi+m)}{N}\exp\left(-\frac{E_{J}^{(m)}}{T}\right),\label{eq:I_eq_w_m-jumps}
\end{equation}
where $Z$ is the corresponding partition function and
\begin{equation}
E_{J}^{(m)}=NJ\left[1-\cos\left(\frac{2\pi(\phi+m)}{N}\right)\left(1-\frac{T}{2J}\right)\right],
\end{equation}
c.f. Eq.$\,$(\ref{E-min}). Harmonic corrections in this formula
are important in the intermediate temperature range for $N\gg1$,
where $m$-fluctuations have to be taken into account but harmonic
approximation still holds.

Equilibrium properties of the system can be computed either by the
Monte Carlo (Metropolis) routine for effective two-component classical
spins $\mathbf{s}_{i}=\left(\sin\theta_{i},\cos\theta_{i}\right)$.
Since at $T\ll J$ equilibration of winding numbers $m$ becomes very
slow, standard Monte Carlo routine using trial changes of directions
of individual spins fails to reach equibrium. However, adding trial
changes of $m$ in the routine,
\begin{equation}
\theta_{i}\rightarrow\theta_{i}^{'}\equiv\theta_{i}+\frac{2\pi m'(i-1)}{N},\qquad0\leq m'\leq N-1
\end{equation}
(one time before or after the full system update by individual rotations)
makes the system equilibrate fast in spite of energy barriers shown
in Fig. \ref{Fig_barrier}.

\begin{figure}[htbp!]
\begin{centering}
\includegraphics[width=8cm]{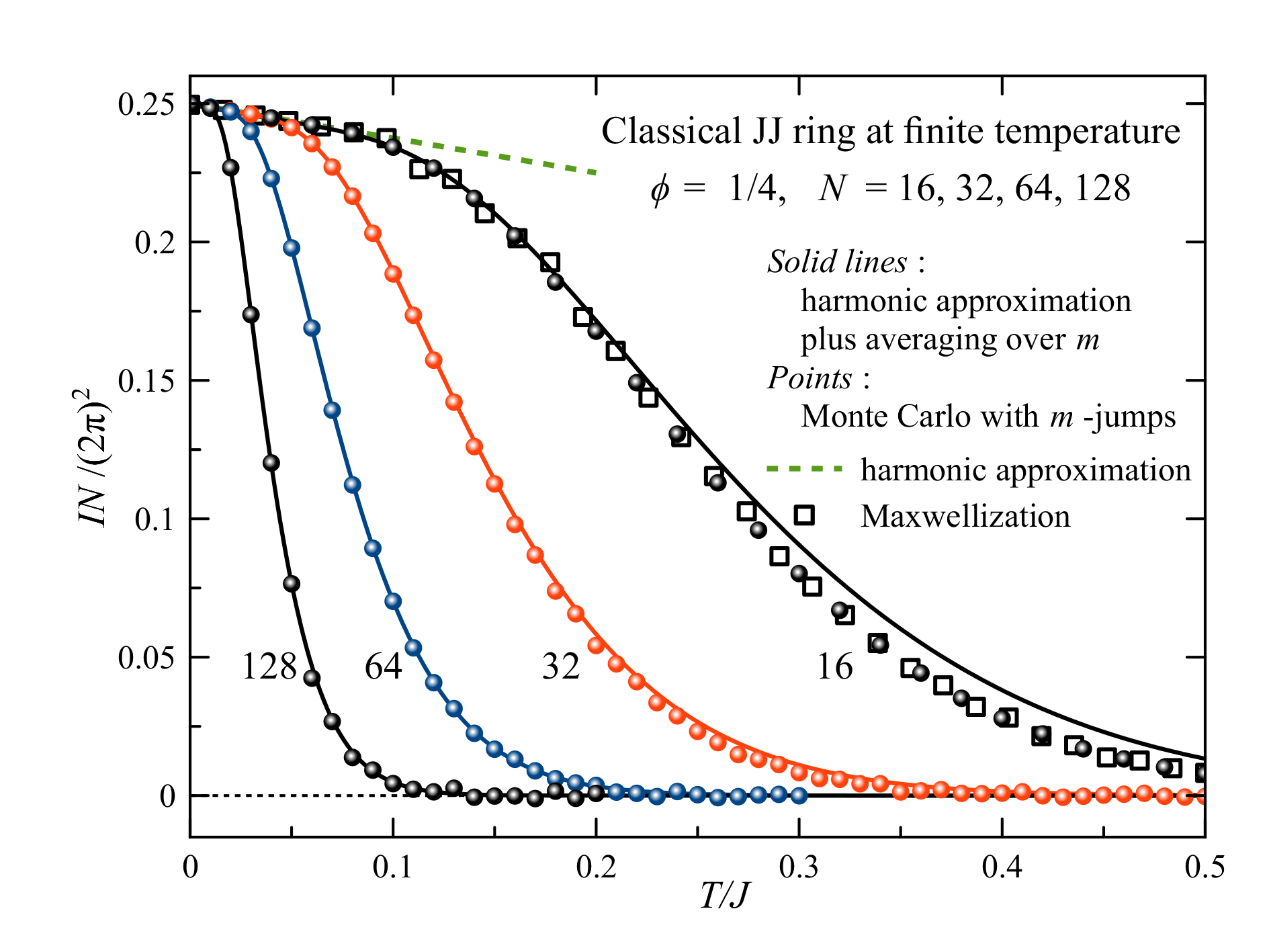}
\par\end{centering}
\caption{Thermal averages of the current at $\phi=1/4$ for different $N$.
Numerical results (symbols) are obtained by Monte Carlo with $m$-jumps
and analytical results (solid lines) are those of Eq.$\,$(\ref{eq:I_eq_w_m-jumps}).
Dashed line is harmonic approximation, Eq.$\,$(\ref{current_harmonic_approx}).}
\label{Fig-I_average-quarter}
\end{figure}

Thermal equilibrium values of the current for different numbers of
grains in the ring are shown in Fig.$\,$\ref{Fig-I_average-quarter},
setting $J=\Phi_{0}=1$. Most of the numerical data were obtained
by Monte Carlo with trial $m$-jumps added, that allows to reach equilibrium
at any temperature. Analytical results of Eq.$\,$(\ref{eq:I_eq_w_m-jumps})
are in accordance with numerical data. In experiment it can be difficult
to reach equilibrium at low temperatures because of energy barriers.
The required equilibration time can be estimated using our dynamical
results below.

\section{Maxwellization method for thermodynamics and equilibrium dynamics }

\begin{figure}[htbp!]
\centering{} \includegraphics[width=8cm]{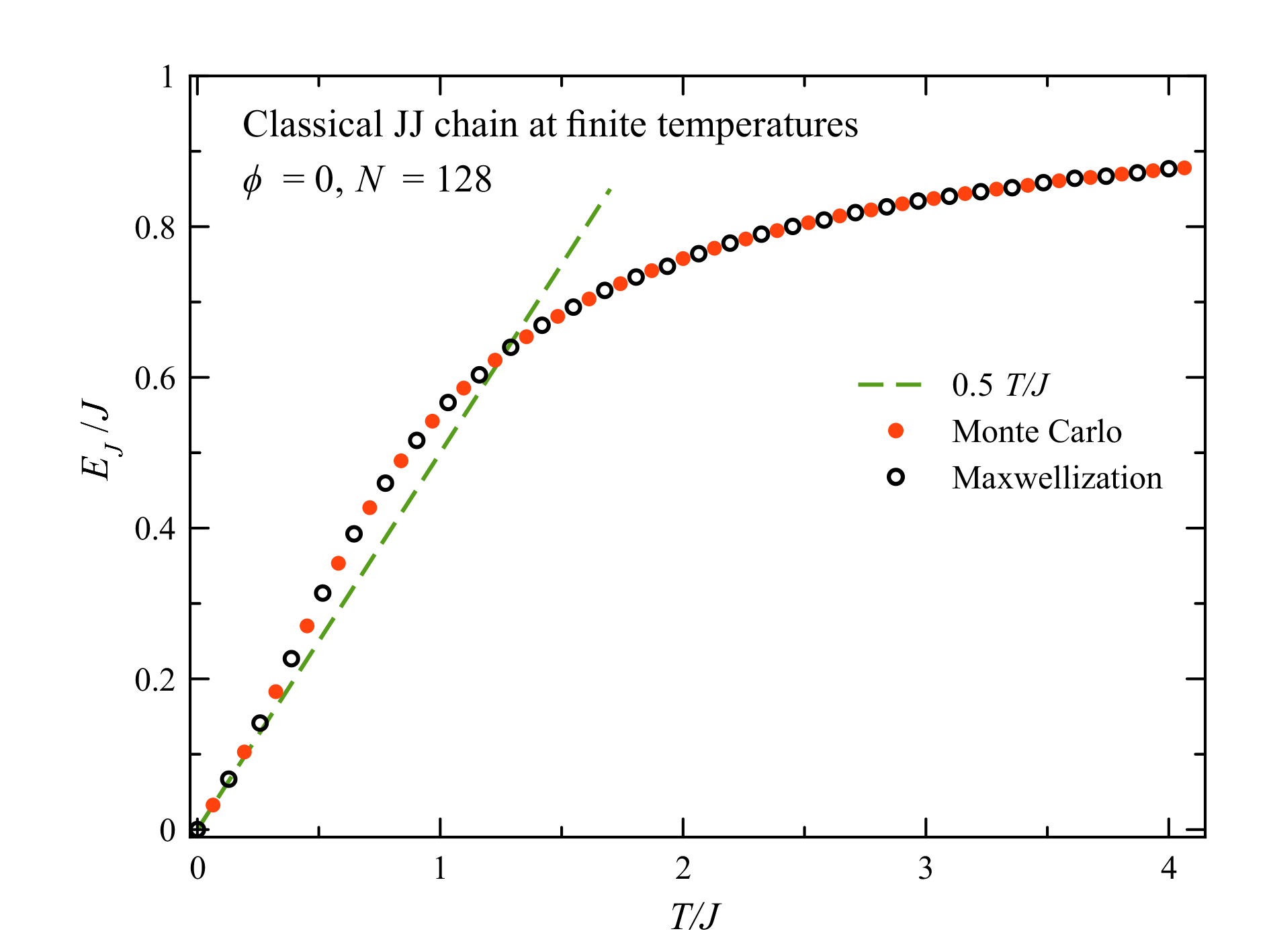} \caption{Josephson energies $E_{J}$ vs $T$, obtained by Monte Carlo and by
maxwellization.}
\label{Fig_MCvsMax}
\end{figure}

While Monte Carlo is a mainstream method at equilibrium, it is not
suitable for dynamical problems simply becuase it is not based on
real dynamics. We propose here another numerical method for statics
and equilibrium dynamics of classical systems having kinetic energy,
such as arrays of Josephson junctions. In this method that we call
\textsl{maxwellization}, equations of motion, here Eq.$\,$(\ref{eq-motion}),
are solved numerically over a long time interval ($0,\tau_{\mathrm{max}})$
divided into sub-intervals of length $\tau_{0}\ll\tau_{\mathrm{max}}$.
At the end of each sub-interval the momenta $p_{i}$, having the Maxwell
distribution $f_{p}\propto\exp\left(-\frac{Jp^{2}}{2T}\right)$ with
the average kinetic energy $T/2$ per particle at equilibrium, are
generated anew with another realization of the Maxwell distribution
at the same temperature $T$, leaving the phases $\theta_{i}$ unchanged.
In such a way a statistical ensemble is created in which the energy
of the system is fluctuating. Kinetic energy is converted into potential
energy during the microscopic time $\tau\sim1$, thus the whole system
becomes quickly thermalized. After a short thermalization time, one
can begin measuring physical quantities by averaging the solution
of the equations of motion over large times. Maxwellization method
works for both equilibrium and dynamic problems.  We have checked
that maxwellization yields the same results with $\tau_{0}\sim1$
and $\tau_{0}\gg1$. Fig.\ \ref{Fig_MCvsMax} shows that thermal
Josephson energies $E_{J}$ obtained by Monte Carlo are the same as
obtained by maxwellization, the accuracy and computer time being comparable.

Maxwellization resuls are also shown in Fig.$\,$\ref{Fig-I_average-quarter}
for $N=16$ with $\tau_{\mathrm{max}}=10^{8}$. For larger $N$ maxwellization
cannot reach equilibrium, similar to standard Monte Carlo without
$m$-jumps. Unlike Monte Carlo, maxwellization cannot be extended
to include $m$-jumps since it is based on realistic dynamics.

\section{Dynamics}

\begin{figure}[htbp!]
\centering \includegraphics[width=8cm]{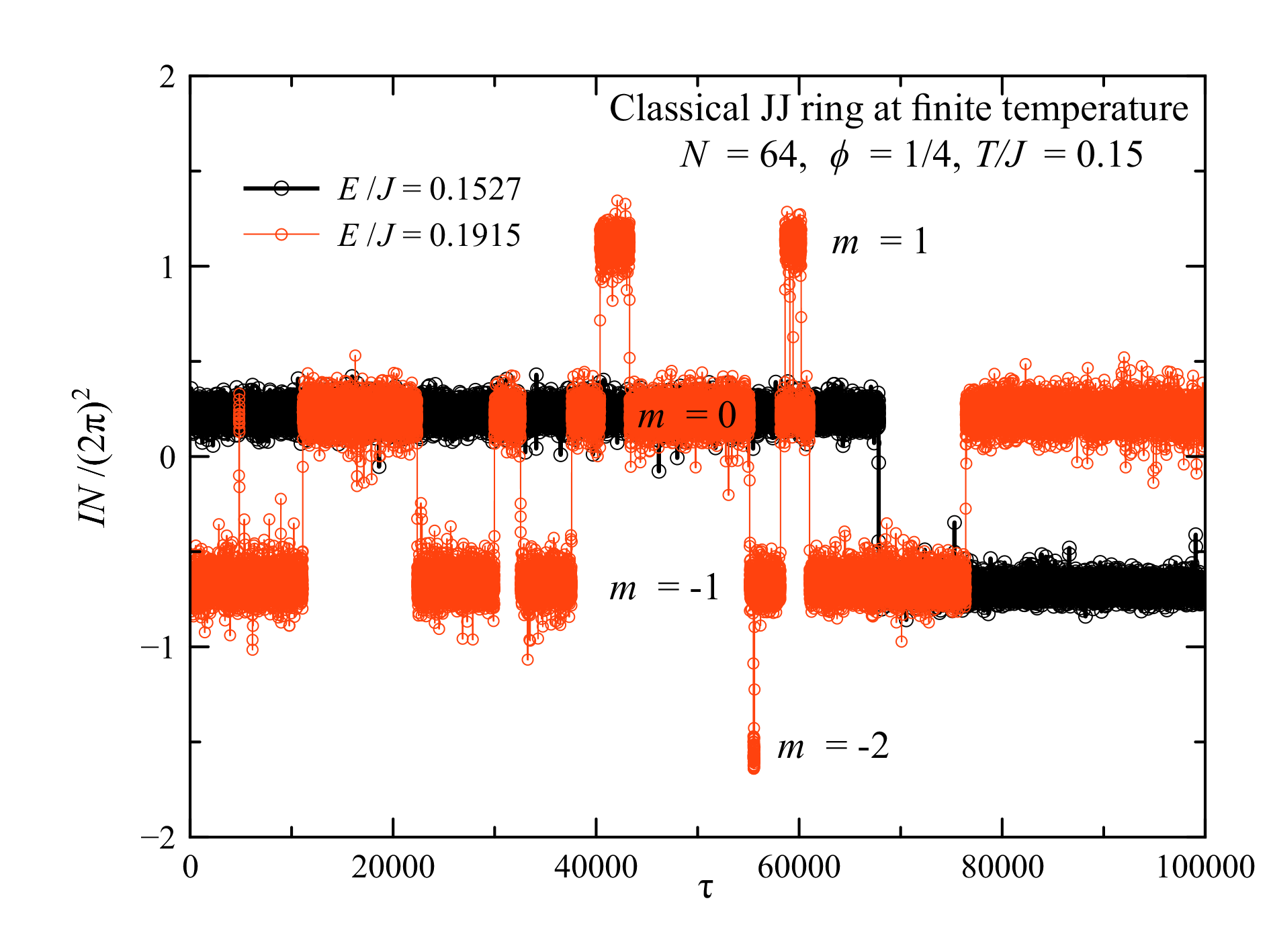} \caption{Time dependence of the current at $\phi=1/4$, showing harmonic fluctuations
and phase slips in states with two different total energies $E$,
generated at the same $T$.}
\label{Fig_I_fluct-quarter}
\end{figure}

Typical time dependences of $I$ obtained by solving the equation
of motion, Eq.$\,$(\ref{eq-motion}), and using Eq.$\,$(\ref{eq:I_Def})
for a quarter-fluxon threading the ring are shown in Fig.\ \ref{Fig_I_fluct-quarter}
in states with two different dynamically conserved energies, generated
at the same temperature. In this illustrative computation, no maxwellization
has been done to show that fluctuations of the current have mainly
dynamic origin. Jumps correspond to the transitions (phase slips)
between different values of $m$ indicated in the figure. Small fluctuations
between phase slips are harmonic excitations. Dynamical fluctuations
of $I$ are stronger in the states with a higher energy. Due to the
lack of symmetry (see Fig.$\,$\ref{Fig_barrier}a), there is a non-zero
time average of the current, shown in Fig.\ \ref{Fig-I_average-quarter}
as the maxwellization result for $N=16$.

\begin{figure}[htbp!]
\centering \includegraphics[width=8cm]{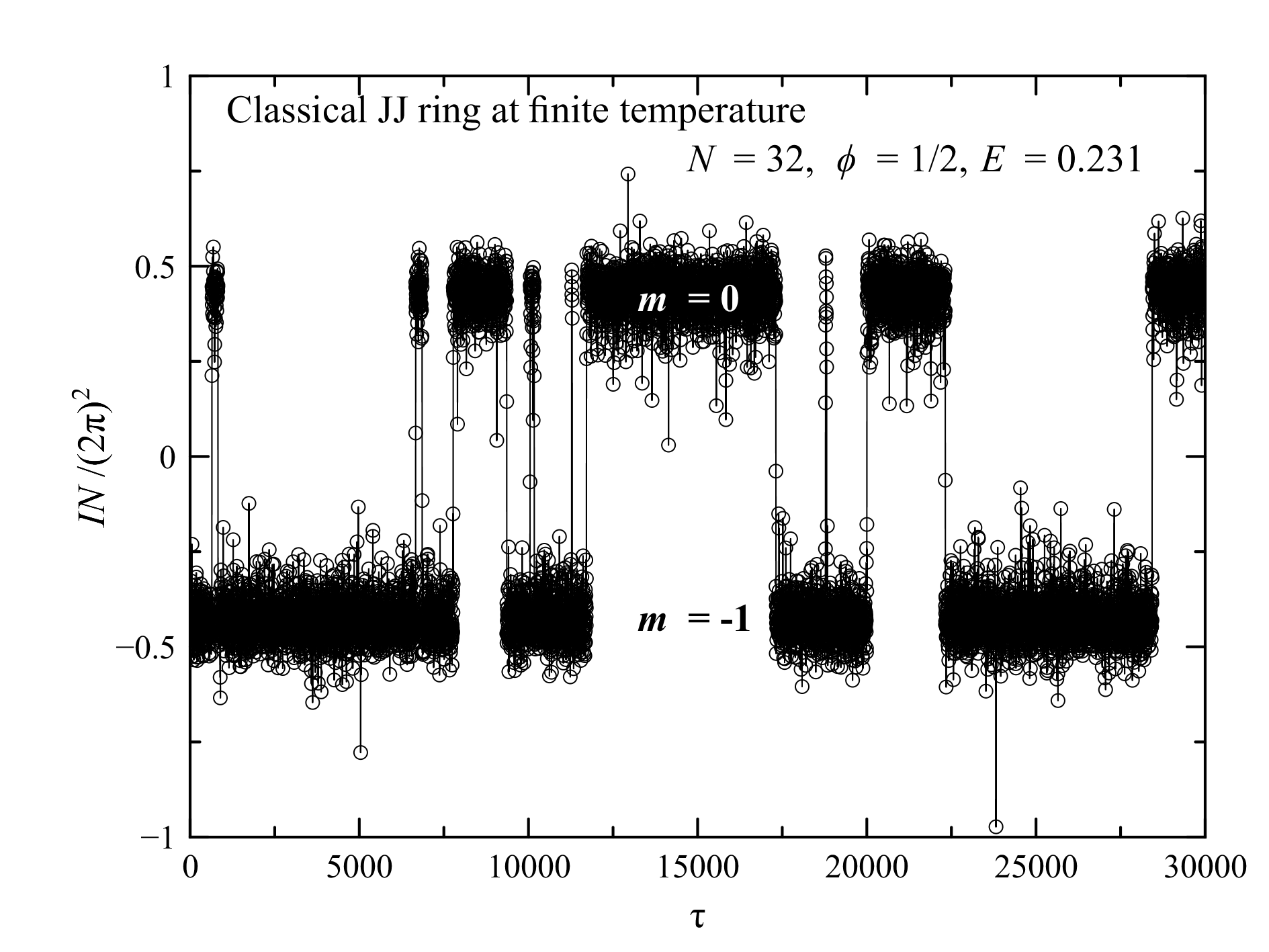} \caption{Time dependence of the Josephson current at $\phi=1/2$, showing harmonic
fluctuations and phase slips.}
\label{Fig_I_fluct-half}
\end{figure}

\begin{figure}[htbp!]
\begin{centering}
\includegraphics[width=8cm]{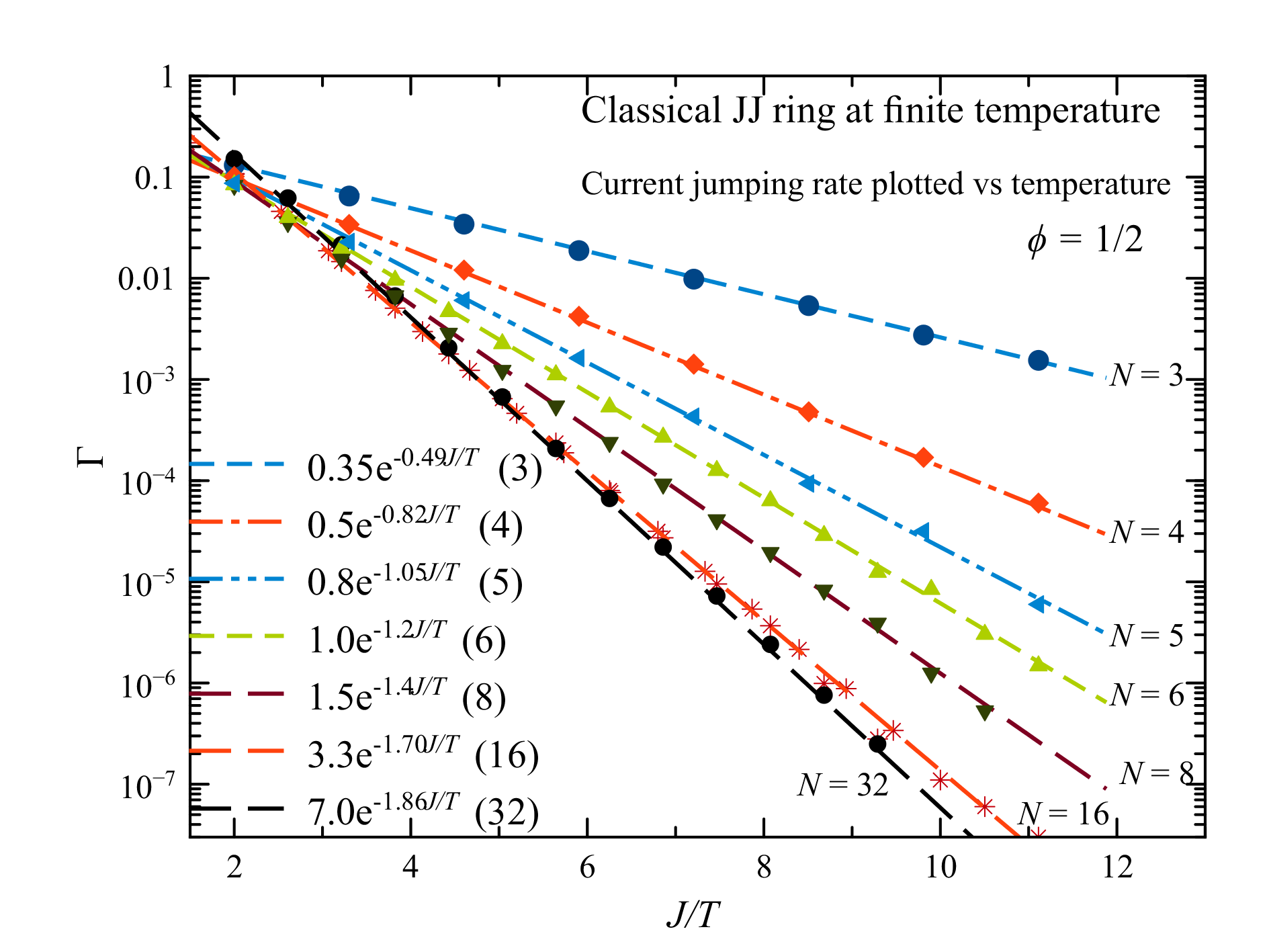}
\par\end{centering}
\caption{Numerically computed transition rates $\Gamma$ with Arrhenius fits
for transitions between opposite directions of the current, corresponding
to $m=0$ and $m=-1$, for rings of different length in the half-fluxon
case. }
\label{Fig_rates}
\end{figure}

\begin{figure}
\begin{centering}
\includegraphics[width=8cm]{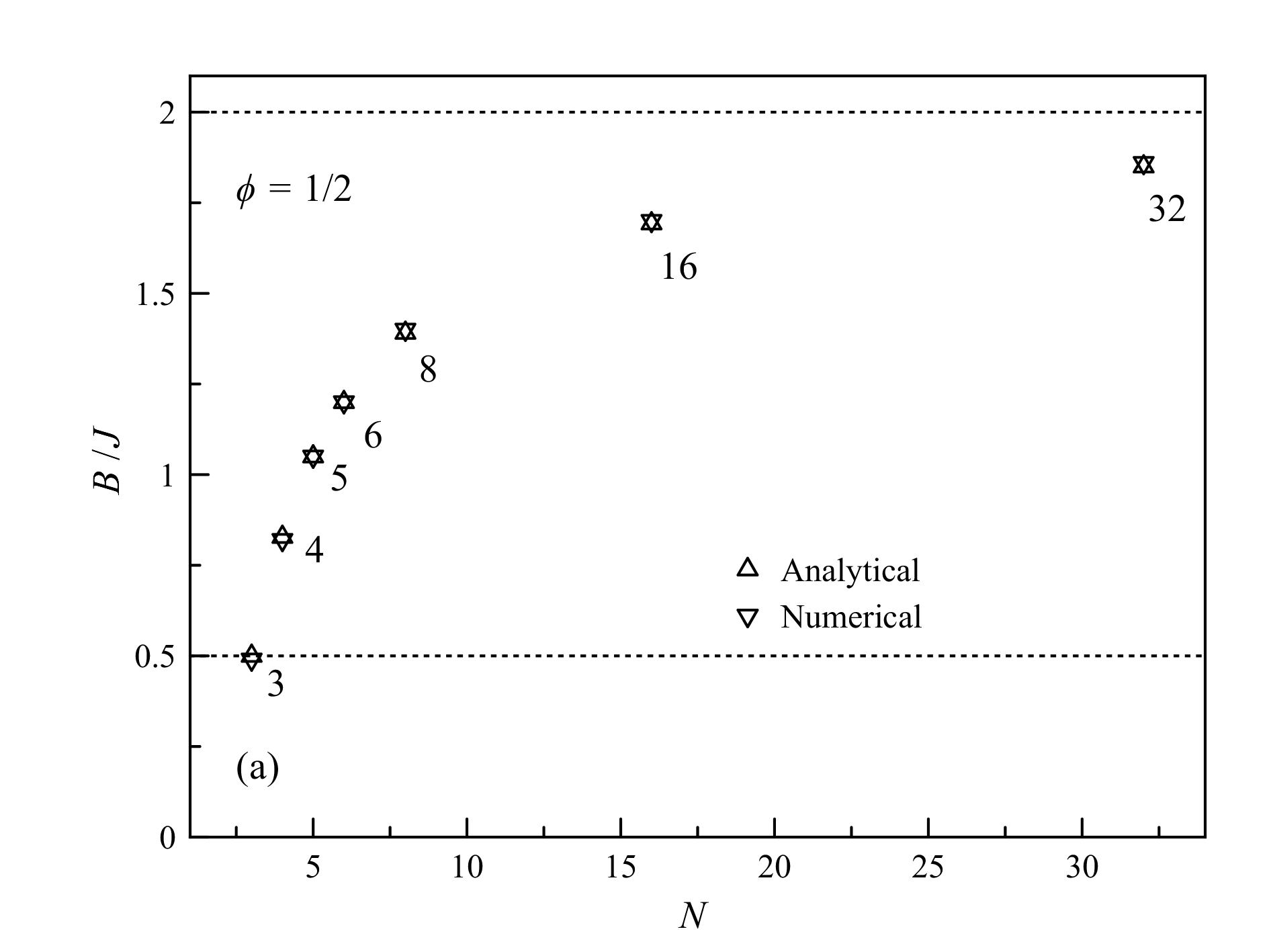}
\par\end{centering}
\begin{centering}
\includegraphics[width=8cm]{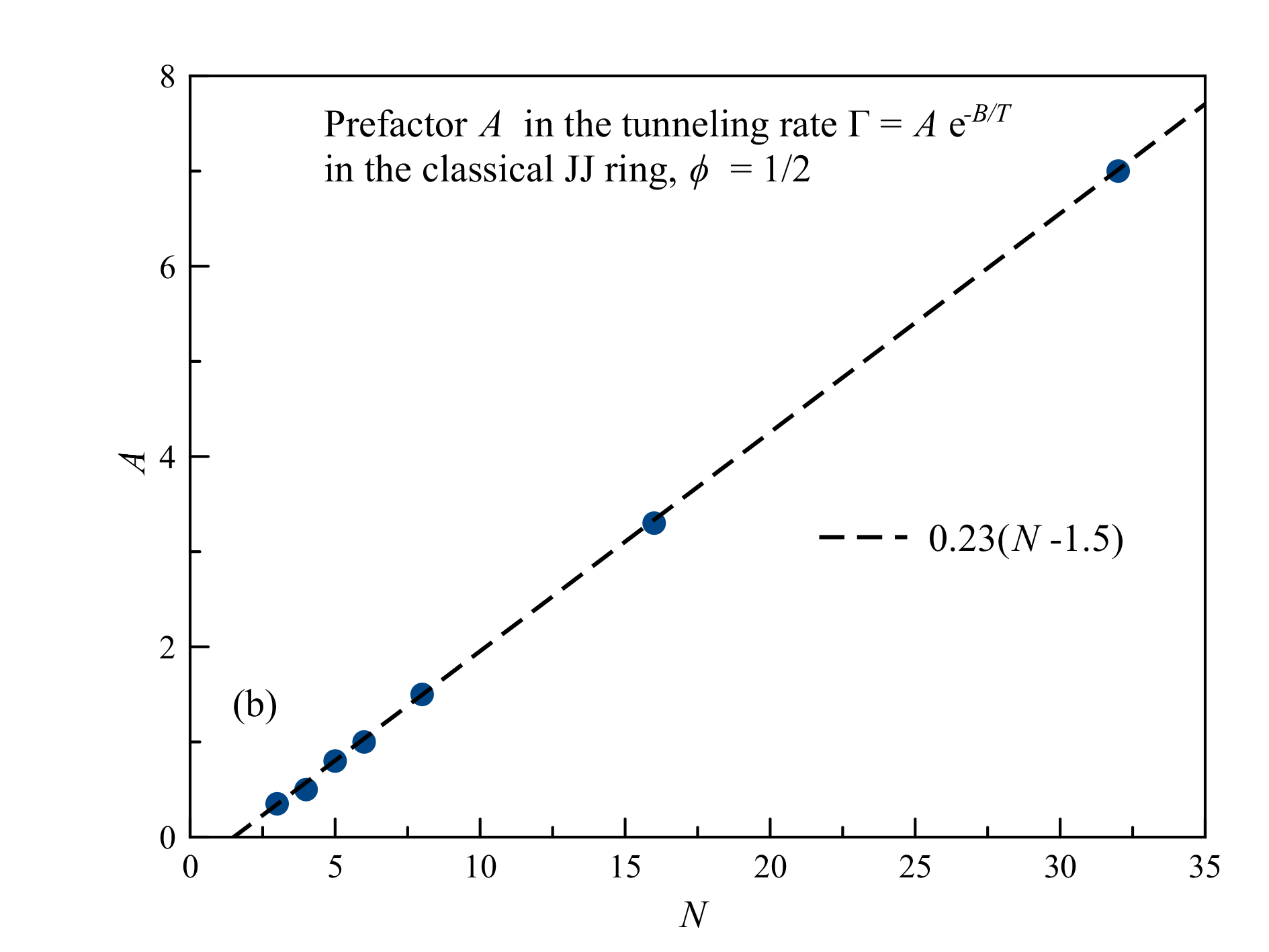}
\par\end{centering}
\caption{Thermal fluctuations of $I$ at a half fluxon. (a) Barrier energies
$B$ extracted from the fits of $\Gamma$ in Fig.$\,$\ref{Fig_rates},
compared to their analytical values from Eq.$\,$(\ref{barrier}).
(b) Prefactor $A$ extracted from numerical data.}

\label{Fig_AB}
\end{figure}

In the half-fluxon case the energy is degenerate and the current averaged
over long times is always zero. $I(t)$ exhibits jumps between opposite
directions corresponding to $m=0$ and $m=-1$, on top of harmonic
fluctuations around these states, see Fig.\ \ref{Fig_I_fluct-half}
(also without maxwellization). The rate of transitions between the
opposite values of $I$ can be computed as $\Gamma=N_{\mathrm{jumps}}/t_{\mathrm{max}}$,
where $N_{\mathrm{jumps}}$ is the number of current jumps within
the time interval of length $t_{\mathrm{max}}$. The results follows
the law $\Gamma=A\exp(-B/T)$, as is shown for closed chains of different
length in Fig.\ \ref{Fig_rates} (in terms of the dimensionless time
$\tau$ with $\tau_{\mathrm{max}}=10^{8}$). Numerically obtained
exponents $B(N)$ are in excellent agreement with the analytical result
given by Eq.\ (\ref{barrier}), as one can see from Fig.$\,$\ref{Fig_AB}.
The computed prefactor is well approximated by $A=0.23(N-3/2)$. Its
proportionality to $N$ at large $N$ agrees with the fact that the
phase slip can occur at any of the $N$ sites of the chain. At higher
temperatures, $T\sim J$, temporal behavior of the current becomes
more chaotic as it involves transitions between other values of $m$
as well.


\section{Discussion}

We have considered equlibrium and dynamic properties of Josephson-junction
rings in the classical limit. It was shown that for rings composed
of many junctions, $N\gg1$, one has to take into account different
values of the winding number $m$ in the thermodynamics of the persistent
current caused by the magnetic flux piercing the ring. Analytical
results combining harmonic approximation with averaging over different
$m$ have been confirmed by a Monte Carlo routine allowing $m$-jumps
(phase slips). Energy barriers for phase slips have been obtained
analytically and shown to increase with $N$.

Numerical method of ``maxwellization'' for solving thermodynamic
and equilibrium dynamic problems has been developed and applied to
JJ rings. This method is based on real dynamics and it is suitable
for computation of quantities such as transition rates at a given
temperature. In particular, we have obtained Arrhenius temperature
dependence of the inversion rate of the persistent current at a half
fluxon with the barrier given by our analytical expressions.

By visualizing the temporal behavior of the current in Josephson junction
chains, our results provide guidance for future experiments in this
field. Similar numerical approach can be tried to study open chains
with a bias current.

\section*{Acknowledgments}

This work has been supported by the grant No. DE-FG02-93ER45487 funded
by the U.S. Department of Energy, Office of Science.


\begin{thebibliography}{10}
\bibitem{Oreg} See, e.g., G. Schwiete and Y. Oreg, Persistent current
in small superconducting rings, Physical Review Letters \textbf{103},
037001-(4) (2009), and references therein.

\bibitem{Mooij-2015} See review and references therein: J. E. Mooij,
G. Schön, A. Shnirman, T. Fuse, C. J. P. M. Harmans, H. Rotzinger,
and A. H. Verbruggen, Superconductor-insulator transition in nanowires
and nanowire arrays, New Journal of Physics \textbf{17}, 033006-(12)
(2015).

\bibitem{Pop} I. M. Pop, I. Protopopov, F. Lecocq, Z. Peng, B. Pannetier,
O. Buisson, and W. Guichard, Measurement of the effect of quantum
phase-slips in a Josephson junction chain, Nature Physics \textbf{6},
589-592 (2010).

\bibitem{Matveev} K. A. Matveev, A. I. Larkin, and L. I. Glazman,
Persistent current in superconducting nanorings, Physical Review Letters
\textbf{89}, 096802-(4) (2002).

\bibitem{Choi-1993} M. Y. Choi, Persistent current and voltage in
a ring of Josephson junctions, Physical Review B \textbf{48}, 15920-15925
(1993).

\bibitem{Bradley} R. M. Bradley and S. Doniach, Quantum fluctuations
in chains of Josephson junctions, Physical Review B \textbf{30}, 1138-1147
(1984).

\bibitem{Wallin-94} M. Wallin, E. S. Sørensen, S. M. Girvin, and
A. P. Young, Superconductor-insulator transition in two-dimensional
dirty boson systems, Physical Review B \textbf{49}, 12115-12139 (1994).

\bibitem{Girvin} See, e.g., S. L. Sondhi, S. M. Girvin, J. P. Carini,
and D. Shahar, Continuous quantum phase transitions, Review of Modern
Physics \textbf{69}, 315-333 (1997).

\bibitem{Sachdev} S. Sachdev, \textit{Quantum Phase Transitions}
(Cambridge University Press, Cambridge, UK, 2011).

\bibitem{Zaikin} A. D. Zaikin, D. S. Golubev, A. van Otterlo, and
G. T. Zimanyi, Quantum phase slips and transport in ultrathin superconducting
wires, Physical Review Letters \textbf{78}, 1552-1555 (1997).

\bibitem{Golubev} D. S. Golubev and A. D. Zaikin, Quantum tunneling
of the order parameter in superconducting nanowires, Physical Review
B \textbf{64}, 014504-(14) (2001).

\bibitem{Korshunov} S. E. Korshunov, Effect of dissipation on the
low-temperature properties of a tunnel-junction chain, Soviet Physics
JETP \textbf{68}, 609-618 (1989).

\bibitem{Chow} E. Chow, P. Delsing, and D. B. Haviland, Length-scale
dependence of the superconductor-to-insulator quantum phase transition
in one dimension. Physical Review Letters \textbf{81}, 204-207 (1998).

\bibitem{Lee-2003} M. Lee, M.-S. Choi, and M. Y. Choi, Quantum phase
transitions and persistent currents in Josephson-junction ladders,
Physical Review B \textbf{68}, 144506-(11) (2003).

\bibitem{garchuPRB16} D. A. Garanin and E. M. Chudnovsky, Quantum
decay of the persistent current in a Josephson junction ring, Physical
Review B \textbf{93}, 094506-(9) (2016).

\bibitem{Rastelli} G. Rastelli, I. M. Pop, and F. W . J. Hekking,
Quantum phase-slips in Josephson junction rings, Physical Review B
\textbf{87}, 174513-(18) (2013).
\end{thebibliography}
\end{document}